\documentclass[aps,twocolumn,showpacs]{revtex4}
\usepackage{amsmath}%
\usepackage{amsthm,amsmath,amssymb}

\usepackage{mathrsfs}
\usepackage{graphicx}% Include figure files
\usepackage{dcolumn}% Align table columns on decimal point
\usepackage{bm}% bold math
\usepackage{color, soul}
\usepackage{color, xcolor}
\usepackage[utf8]{inputenc}
\usepackage[T1]{fontenc}
\usepackage{mathptmx}
\usepackage{multirow}
\usepackage{makecell}
\usepackage{setspace}
\usepackage{footnote}
\usepackage{hyperref}
\hypersetup{hypertex=true,
	colorlinks=true,
	linkcolor=blue,
	anchorcolor=blue,
	citecolor=blue}

\begin{document}
\title{Dark solitons and their bound states in a nonlinear fiber with second- and fourth-order dispersion}
%\title{Stripe optical solitons in the competition between the second- and fourth-order dispersion}

%\author{XXX$^{1}$}
\author{Peng Gao$^{1}$}\email{pgao@gscaep.ac.cn}
\author{Li-Zheng Lv$^{2}$}%\email{lizhenglv@stu.xjtu.edu.cn}
\author{Xin Li$^{3}$}%\email{lixin_2021@bupt.edu.cn}

\address{$^1$Graduate School, China Academy of Engineering Physics, Beijing 100193, China}
\address{$^2$School of Physics, Xi’an Jiaotong University, Xi’an 710049, China}
\address{$^3$School of Science, Beijing University of Posts and Telecommunications, Beijing 100876, China}

 %%%%%%%%%%%%%%%%%%%%%%%%%%%%%%%%%%%%%%%%%%%%%%%%%
%\date{Mar. 1, 2018}
\begin{abstract}
We study the excitations of dark solitons in a nonlinear optical fiber with the second- and fourth-order dispersion, and find the emergence of striped dark solitons (SDSs) and some multi-dark-soliton bound states.
The SDSs can exhibit time-domain oscillating structures on a plane wave, and they have two types: the ones with or without the total phase step, while the multi-dark-soliton bound states exhibit different numbers of amplitude humps.
By the modified linear stability analysis, we regard the SDSs as the results of the competition between periodicity and localization, and analytically give their existence condition, oscillation frequency, and propagation stability, which show good agreements with numerical results.
We also provide a possible interpretation of the formation of the existing striped bright solitons (SBSs), and find that SBS will become the pure-quartic soliton when its periodicity and localization keep balance.
Our results provide the theoretical support for the experimental observation of striped solitons in nonlinear fibers, and our method can also guide the discovery of striped solitons in other physical systems.
\end{abstract}

\maketitle

\section{Introduction}

Periodicity and localization are two fundamental but distinct  characteristics of nonlinear waves.
The periodicity is a commonality of all waves and dominates various wave phenomena \cite{Hecht-book,Cronin-2009}, while the (self-induced) localization dominates the formation of nonlinear localized waves \cite{Kivshar-book,Onorato-2013,Dudley-2014}.
As a widespread kind of nonlinear localized wave, soliton has been attracting lots of attentions and has wide applications by virtue of its property of stable propagation \cite{Mollenauer-1980,Weiner-1988,Agrawal-book,Kevrekidis-book,Haus-1996,McDonald-2014}.
Soliton can produce periodicity-dominated phenomena like interference \cite{Snyder-1997,Kumar-2009,Zhao-2016} and diffraction \cite{Xiong-2023,Gao-2023}, and it also has localization-dominated phenomena like self-bound propagation and elastic collision \cite{Zabusky-1965,Stegeman-1999}.
In the previous studies, the periodicity and localization of a soliton were always treated separately, and little attention has been paid to their combined effects.
However, in nonlinear fibers with the fourth-order dispersion (FOD), the bright solitons with oscillating tails \cite{Akhmediev-1994,Buryak-1995,Blanco-2016,Tam-2019} and even stripe structures \cite{Zakharov-1998,Tam-2020} were discovered.
They exhibit both of noticeable periodicity and localization, and therefore can be considered as the results of competition between the two characteristics.
As we know, the modified linear stability analysis (MLSA) method has been used to predict the quantitative dynamics of nonlinear waves, including their periodicity and localization  \cite{Gao-2020,Gao-2021}.
Therefore, it provides the possibility to figure out the formation mechanism of striped solitons under the FOD and furthermore find more kinds of solitons in this system.

In this paper, we study the excitations of dark solitons in a nonlinear optical fiber with the second-order dispersion and FOD.
We find that striped dark solitons (SDSs) and their bound states exist in this system, which exhibit more kinds of structures than the striped bright solitons \cite{Zakharov-1998,Tam-2020} and the multi-soliton bound states \cite{Buryak-1995}.
The single-soliton state of SDS contains two types, i.e., the ones with or without the total phase step, while the multi-dark-soliton bound states have different numbers of amplitude humps in the time-domain distribution.
By the MLSA method, a SDS is considered as the result of the competition between periodicity and localization, and furthermore we quantitatively give its existence condition and oscillation frequency.
Based on the derived oscillation frequency, we analyze the modulation instability (MI) of plane-wave background of SDS and give its stability condition.
Besides, we apply the MLSA to the SBSs to provide a possible interpretation of their formation, and find that the pure-quartic soliton is a special kind of SBS when its periodicity and localization keep balance.

\section{Model and the existence condition of striped solitons}

In the slow-varying envelope approximation, the propagation of a linearly polarized light in a single-mode nonlinear fiber can be described by the following nonlinear Schr\"{o}dinger equation with the second-order dispersion and FOD \cite{Agrawal-book},
\begin{align}\label{eq-modelexp}
    i\frac{\partial A}{\partial Z}-\frac{\beta'_2}{2}\frac{\partial^2 A}{\partial T^2}+\frac{\beta'_4}{24}\frac{\partial^4 A}{\partial T^4}+\gamma|A|^2A=0,
\end{align}
where $A(T,Z)$ represents the slowly varying complex envelope of optical field, and $Z$, $T$ are the evolution distance and retarded time.
$\beta'_2$, $\beta'_4$, and $\gamma$ are the coefficients of second-order dispersion, FOD, and Kerr nonlinearity, respectively.
By the transformation $A=\sqrt{P_0}\psi$, $Z=U_z z$, and $T=U_t t$, the model (\ref{eq-modelexp}) can be transformed into a dimensionless model,
\begin{align}\label{eq-model}
    i\frac{\partial \psi}{\partial z}-\frac{\beta_2}{2}\frac{\partial^2 \psi}{\partial t^2}+\frac{\beta_4}{24}\frac{\partial^4 \psi}{\partial t^4}+|\psi|^2\psi=0,
\end{align}
where $\beta_2=\beta'_2 U_z/U_t^2$, $\beta_4=\beta'_4 U_z/U_t^4$, and $P_0$ is the background value of the incident light power.
The two quantities, $U_z$ and $U_t$, respectively have the units of space and time.
In Ref. \cite{Runge-2020}, to generate a pure-quartic soliton in a fiber laser, the parameters were set as $\beta'_2=21.4\,{\rm ps^2/km}$, $\beta'_4=-80\,{\rm ps^4/km}$, $\gamma=1.6\,{\rm /W/km}$, and $P_0=0.37\,{\rm W}$. 
When setting $U_z=1/\gamma P_0=1.69\,{\rm km}$ and $U_t=6\,{\rm ps}$, one can obtain the dimensionless parameters, $\beta_2=1$ and $\beta_4=-0.13$.
In our work, the range of $\beta_2$ and $\beta_4$ will be extended to generate more kinds of solitons.

\begin{figure}[htbp]
	\centering
	\includegraphics[width=80mm]{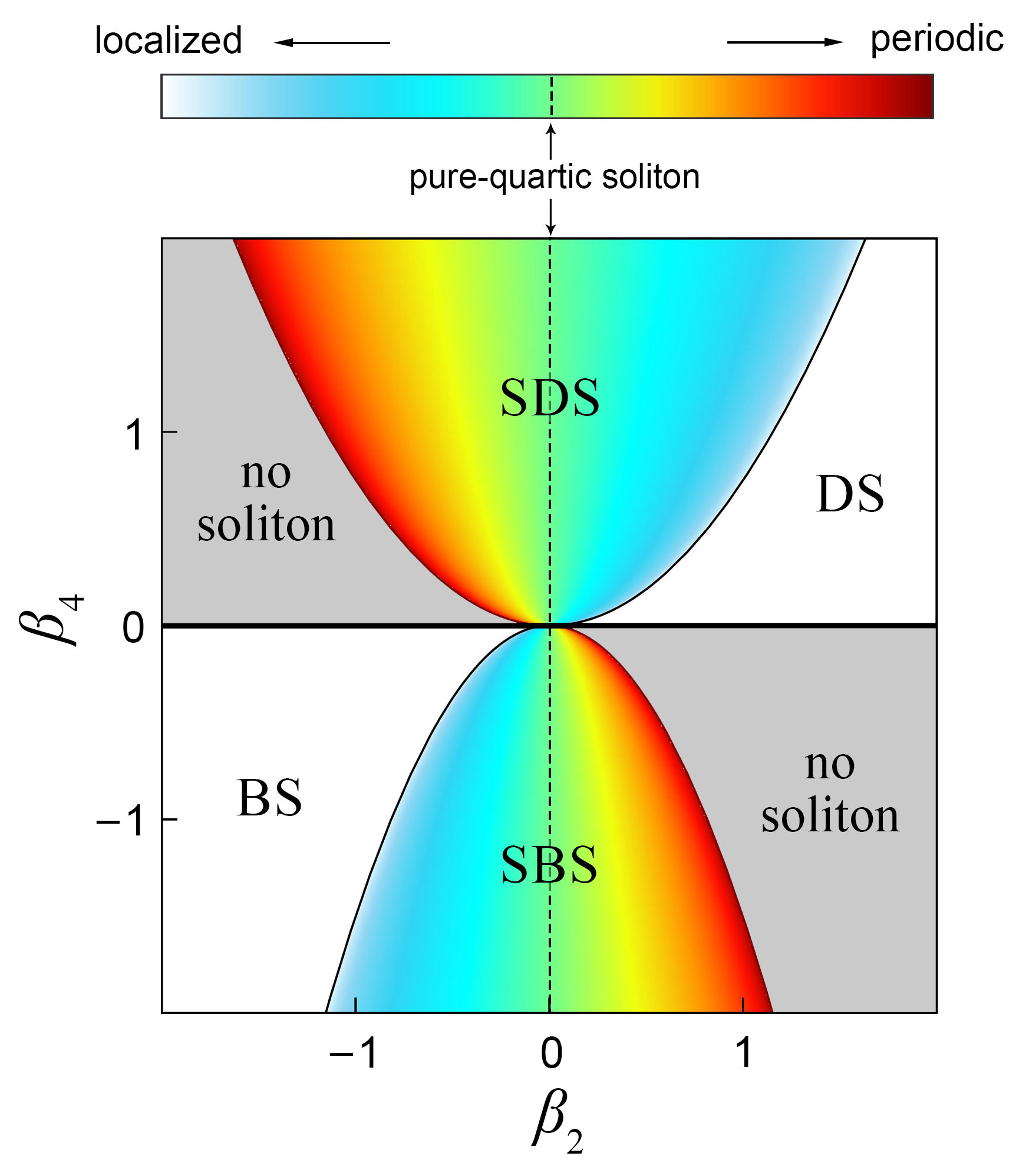}
	\caption{(Color online) Existence condition of solitons under different second-order dispersion $\beta_2$ and FOD $\beta_4$.
		They contain the regions of SDS (striped dark soliton), DS (the traditional dark soliton), SBS (the striped bright soliton), and BS (the traditional bright soliton).
		The boundary of SDS region has the expression $\beta_4=3\beta_2^2/4\beta$, and the boundary of SBS region is $\beta_4=-3\beta_2^2/2\beta$, where $\beta$ is the propagation constant of solitons.
		The color scale describes the strength $\varphi$ of a soliton's periodicity relative to its localization in the range from 0 to $\pi/2$.
		The black dashed line denotes the case of $\beta_2=0$, where a pure-quartic soliton can exist.
		In all cases, the propagation constant is set as $\beta=1$.
	}
	\label{pic-ph}
\end{figure}

In the model (\ref{eq-model}), some solitons and multi-soliton states have been studied \cite{Karlsson-1994,Piche-1996,Akhmediev-1994,Buryak-1995,Blanco-2016,Tam-2019,Zakharov-1998,Tam-2020}, including their existence conditions, stability, and even analytical solutions. 
All of these solitons belong to bright solitons, which have no plane waves as their background waves. 
In our work, the solitons we mainly concern about are the dark solitons, which exist on the plane-wave background, and therefore are essentially different from the bright solitons.
According to our analysis, the existence condition of solitons in the model (\ref{eq-model}) is shown in Fig. \ref{pic-ph}.
The solitons with different structures correspond to different regions in the $\beta_2$-$\beta_4$ plane. The details are as follows.
\begin{itemize}
    \item The horizontal line: when $\beta_4=0$, the model (\ref{eq-model}) becomes the standard nonlinear Sch\"{o}dinger model, where only the traditional dark ($\beta_2>0$) or bright ($\beta_2<0$) solitons can exist.
    \item The region of SDS: when $\beta_4>{3\beta_2^2}/{4\beta}$, the SDSs exist and their periodicity gets stronger with $\beta_2$ decreasing [see Eq. (\ref{eq-ec-sds})].
    \item The region of DS: when $\beta_4\leq{3\beta_2^2}/{4\beta}$ and $\beta_2>0$, the traditional dark soliton exists, which has no stripe.
    \item The region of SBS: when $\beta_4<-{3\beta_2^2}/{2\beta}$, the SBSs exist and their periodicity gets stronger with $\beta_2$ increasing [see Eq. (\ref{eq-ec-sbs})]. 
    The solitons presented in Refs. \cite{Akhmediev-1994,Buryak-1995,Blanco-2016,Tam-2019,Zakharov-1998,Tam-2020} are in this region. In particular, it becomes the so-called pure-quartic soliton when $\beta_2=0$ \cite{Blanco-2016}.
    \item The region of BS: when $\beta_4<-{3\beta_2^2}/{2\beta}$ and $\beta_2<0$, the traditional bright soliton exists, which has no stripe. The solitons presented in Refs. \cite{Karlsson-1994,Piche-1996} are in this region.
\end{itemize}
In the above conditions, $\beta(>0)$ denotes the propagation constant of solitons.
In other regions of the $\beta_2$-$\beta_4$ plane, no soliton exists.
We define the quantity $\varphi$ by $\varphi=\tan^{-1} ({\omega_s}/{\eta_s})$ in the range from $0$ to $\pi/2$ to describe the strength of a wave's stripe characteristic.
$\omega_s$ and $\eta_s$ are respectively the parameters determining the periodicity and localization of a wave, whose expressions can be seen in Eqs. (\ref{eq-omet-sds}) and (\ref{eq-omet2}) for SDS and SBS, respectively.
With $\varphi$ increasing, the wave's periodicity gets stronger and the wave's localization gets weaker.
The striped solitons with different shapes can be regarded as the results of the competition between periodicity and localization in different proportions.
In particular, for a pure-quartic soliton \cite{Blanco-2016}, the two parameters $\omega_s$ and $\eta_s$ are equal, which indicates that its periodicity and localization keep balance.

In the next two sections, we will separately study SDS and SBS, including their existence condition, stripe characteristic, and formation mechanism.

\begin{figure*}[htbp]
\centering
\includegraphics[width=176mm]{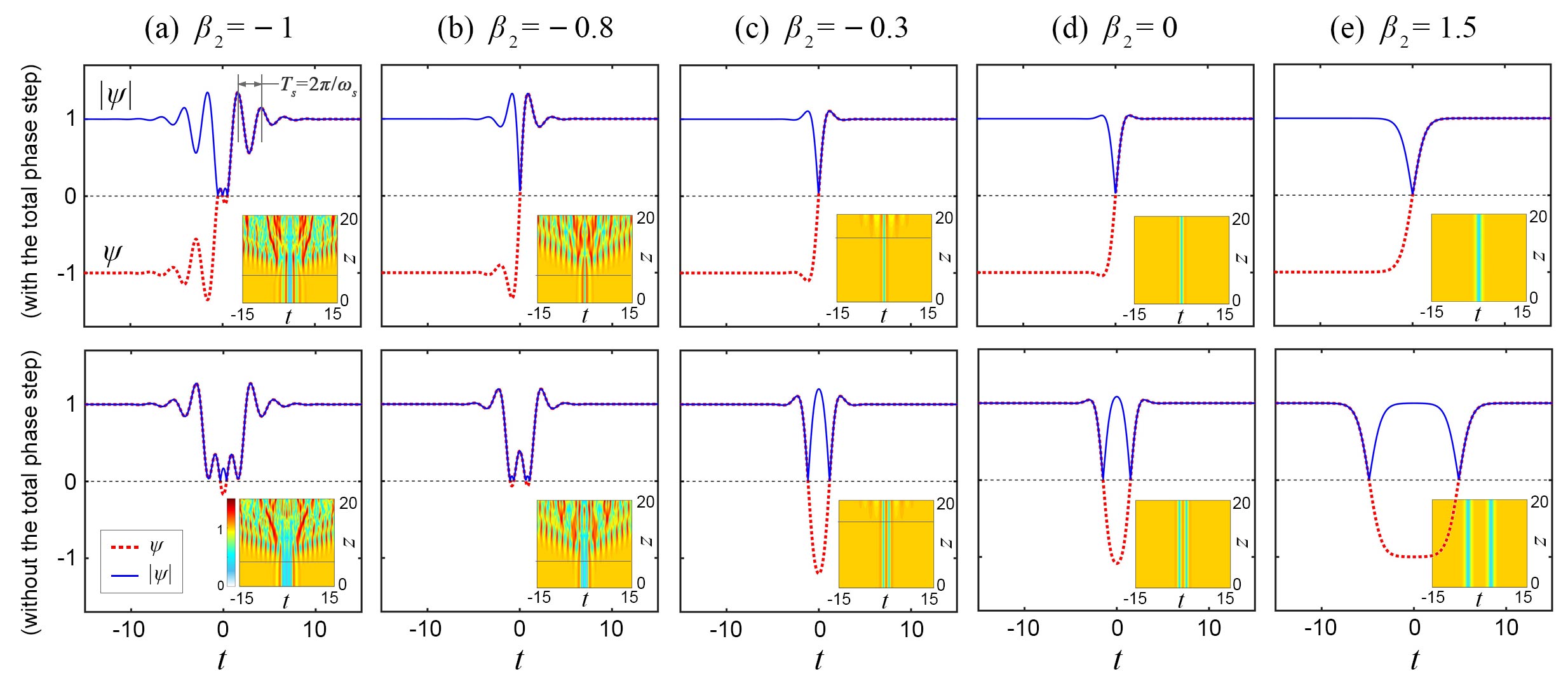}
\caption{(Color online) The dark solitons with (upper row) or without (lower row) the total phase step.  Distributions of the wave functions $\psi$ of dark solitons (red dashed curves) and their amplitude $|\psi|$ (blue solid curves), when the second-order dispersion $\beta_2$ is (a) $-1$, (b) $-0.8$, (c) $-0.3$, (d) 0, and (e) 1.5.
The insets are their corresponding amplitude evolution plots, where the horizontal lines are the spatial dividers between the solitons and the spontaneous oscillations induced by MI.
Other parameters are set as $\beta_4=1$ and $\beta=1$.
}
\label{pic-ds}
\end{figure*}

\section{SDS}

\subsection{Single-soliton state}

At first, let us focus on the single-soliton states of SDSs and their counterparts without the total phase step.
The term ``the total phase step'' is used to describe the phase difference of plane wave between the two sides of dark solitons, which is the key characteristic of dark solitons.
It is known that the model (\ref{eq-model}) has the plane-wave solution,
\begin{align}\label{eq-pw}
	\psi_0=a_0e^{i\beta z},
\end{align}	
where the amplitude $a_0$ and the propagation constant $\beta$ have the relationship $\beta=a_0^2$.
In this section, we set $a_0=1$ so that $\beta=1$.
Considering that it is quite difficult to analytically solve Eq. (\ref{eq-model}), we numerically solve it by the biconjugate gradient method \cite{Press-book}.
This numerical method has been successfully applied to obtain dark solitons or vertices in different models of Bose-Einstein condensates \cite{Winiecki-1999,Edmonds-2016}.
By the method, when $\beta_4=1$, we obtain the numerical solutions of dark solitons under different $\beta_2$ and show them in Fig. \ref{pic-ds}.
When $\beta_2=-1$, we find two kinds of fundamental SDSs, which has or does not have the total phase step (i.e., the key characteristic of dark solitons), and we show them at the top or bottom of Fig. \ref{pic-ds} (a), respectively.
Both of them have some stripes and oscillating tails, but their structures are different.

Later, we set their wave functions as the initial states to numerically integrate the model (\ref{eq-model}), by the split-step Fourier method \cite{Yang-book}.
The evolution results are shown as the insets in Fig. \ref{pic-ds} (a).
The SDS with the total phase step stably propagates for seven unit distance.
At its final evolution stage, some new waves can be seen.
Because the plane-wave background is unstable under the perturbation of dark solitons, these waves are generally induced by the modulation instability (MI), which will be analyzed in detail below.
The similar result is also obtained for the SDS without the total phase step, which propagates stably for six unit distance.

For some larger values of $\beta_2$, the results are shown in Figs. \ref{pic-ds} (b), (c), (d), and (e).
With the increase of $\beta_2$, the SDS has less stripes but can propagate for a longer distance, especially when $\beta_2>0$.
Note that the SDS without the total phase step can be regarded as two separated dark solitons when $\beta_2=1.5$, as shown in Fig. \ref{pic-ds} (e).
It indicates that, when the SDS does not exist, the dark soliton without the total phase step does not exist.

\subsection{Analysis of existence condition and oscillation frequency}

\begin{figure}[htbp]
	\centering
	\includegraphics[width=86mm]{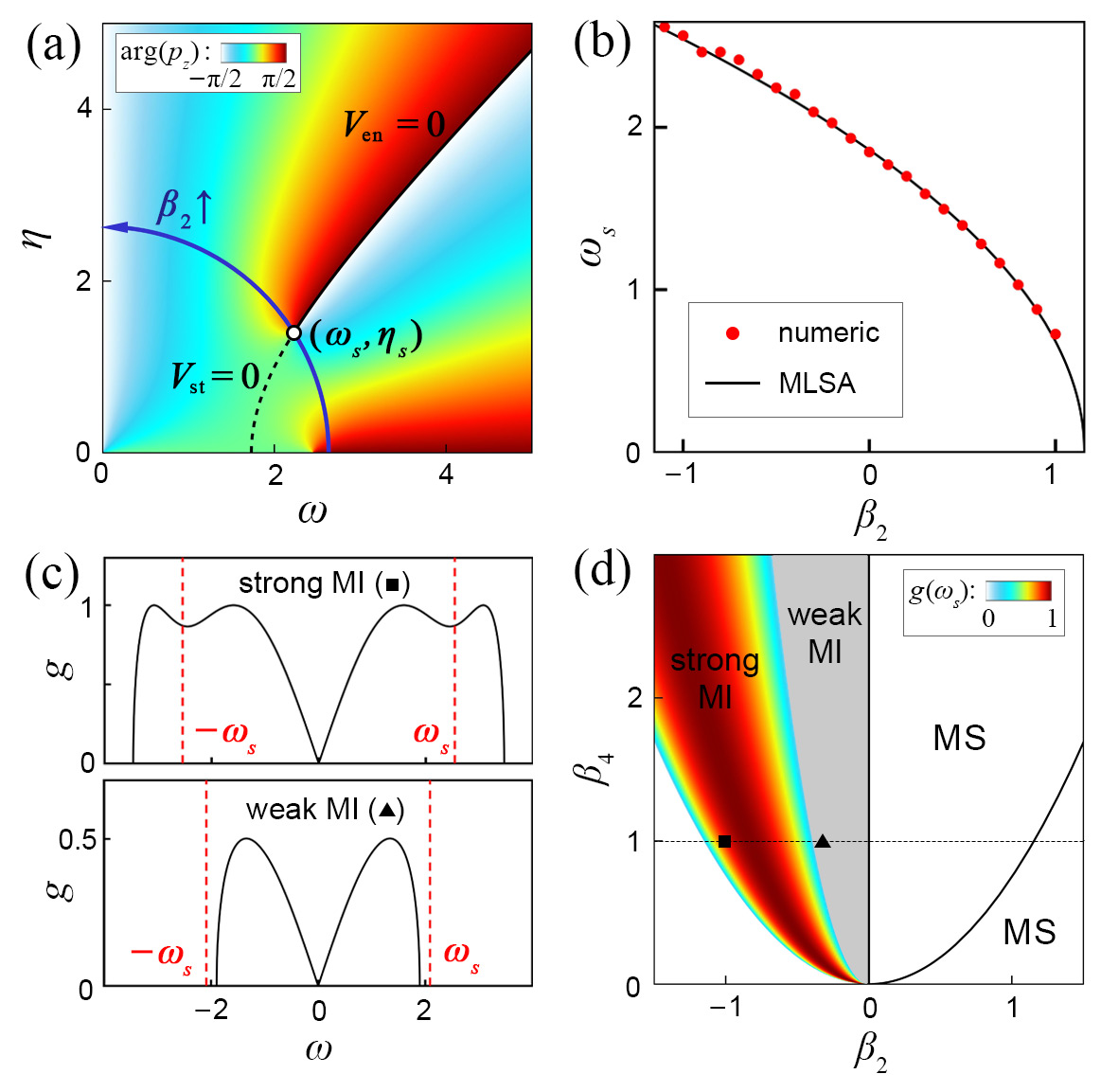}
	\caption{(Color online) (a) Phase of the propagating function $p_z(t,z)$ under different $\omega$ and $\eta$, when $\beta_2=-0.5$, $\beta_4=1$, and $\beta=1$. The black solid and dashed curves denote $V_{\rm en}=0$ and $V_{\rm st}=0$, respectively. The white point with the coordinate $(\omega_s, \eta_s)$ is an exceptional point of $\arg(p_z)$ function. The yellow curve with an arrow is the trajectory of this point with $\beta_2$ increasing.
		(b) Dependence of SDS's oscillation frequency $\omega_s$ on $\beta_2$, when $\beta_4=1$. The red dots and the black curve are respectively the results from numerical evolution and the MLSA method.
		(c) Dependence of the gain value $G$ on $\omega$, when $\beta_2=-1$ (upper) or $\beta_2=-0.3$ (lower). The strong or weak MI is defined by whether $\omega_s$ locates in the MI band or not.
		(d) Distribution of MI (modulation instability) and MS (modulation stability) on the $\beta_2$-$\beta_4$ plane. The color scale describes the value of $g(\omega_s)$, i.e., Eq. (\ref{eq-goms}).
	}
	\label{pic-mi}
\end{figure}

For the bright solitons in the model (\ref{eq-model}), the asymptotic analysis has been used to obtain their existence condition and oscillation characteristic \cite{Akhmediev-1994,Buryak-1995,Tam-2020,Zakharov-1998}.
However, considering that the dark solitons have plane-wave backgrounds, the asymptotic analysis cannot be applied directly to them.
As shown in Fig. \ref{pic-ds}, SDSs have the background waves, and meanwhile they have both of periodicity and localization.
For such a nonlinear wave, the MLSA method can quantitatively give its dynamical characteristics and guide its generation \cite{Gao-2020,Gao-2021}.
Thus, we will apply this method to analyze the existence condition and characteristics of fundamental SDSs.
It starts with the ansatz solution of a perturbed plane wave,
\begin{align}\label{eq-psip}
    \psi_p=\psi_0(1+u),
\end{align}
where the plane wave $\psi_0$ has the expression (\ref{eq-pw}), and $u$ denotes a perturbation exerted on it.
Substituting Eq. (\ref{eq-psip}) into the model (\ref{eq-model}) and linearizing it about $u$, we can obtain
\begin{align}\label{eq-modelu}
    i\frac{\partial u}{\partial z}-\frac{\beta_2}{2}\frac{\partial^2 u}{\partial t^2}+\frac{\beta_4}{24}\frac{\partial^4 u}{\partial t^4}+\beta(u+u^*)=0,
\end{align}
where $u^*$ denotes the complex conjugation of $u$, and we have $\beta=a_0^2$.
If the ansatz $u=e^{\lambda t}$ is applied to Eq. (\ref{eq-modelu}) according to Refs. \cite{Akhmediev-1994,Tam-2020,Zakharov-1998}, the term $u^*$ in Eq. (\ref{eq-modelu}) will be neglected and the results of dark solitons will be the same as the ones of bright solitons.
It will cause that the derived existence region and oscillation frequency of SDSs don't agree with the numerical results of Figs. \ref{pic-ds} and \ref{pic-mi} (b).

Therefore, to take the term $u^*$ into consideration, we assume the perturbation as the linear superposition of two conjugating components,
\begin{align}\label{eq-uab}
    u=Ae^{p(t,z)}+Be^{p^*(t,z)},
\end{align}
where $A$ and $B$ are the amplitudes of the two components, and $p(t,z)$ is a complex function.
This ansatz is a key setting of the MLSA method.
Substituting Eq. (\ref{eq-uab}) into the perturbation's model (\ref{eq-modelu}) and only considering the parts of $e^{p(t,z)}$, one can obtain two linear equations of $A$ and $B^*$.
They can be written as the following form,
\begin{align}\label{eq-abc}
	i\begin{bmatrix}
	-M+\beta & \beta \\
	-\beta & M-\beta
	\end{bmatrix}\begin{bmatrix}
	A \\ B^*
	\end{bmatrix}=p_z\begin{bmatrix}
	A \\ B^*
	\end{bmatrix},
\end{align}
where
\begin{align}
M=&\frac{\beta_2}{2}p_{tt}-\frac{\beta_4}{24}p_t^4-\frac{\beta_4}{8}p_{tt}^2-\frac{\beta_4}{6}p_tp_{ttt}\nonumber\\
&-\frac{\beta_4}{24}p_{tttt}-p_t^2(-\frac{\beta_2}{2}+\frac{\beta_4}{4}p_{tt}).
\end{align}
In the above expressions, $p_t$ and $p_z$ denote the partial derivative of $p(t,z)$ about $t$ and $z$, respectively.
The function $p_z$ can be obtained by solving the eigenvalue of the left-hand matrix in Eq. (\ref{eq-abc}):
\begin{align}
    p_z=&\pm\sqrt{M(2\beta-M)}.
\end{align}
There are two modes, where the eigenvalues $p_z$ are the opposite of each other.
The mode with a positive eigenvalue is focused on here.

Now, our main aim is to find the condition that the shape of the perturbation remains unchanged in the propagating process, which is just the existence condition of SDS.
We consider the function $p(t,z)$ has the following form at the initial distance,
\begin{align}\label{eq-p}
    p(t,z=0)=i\omega t+\ln [{\rm sech}(\eta t)],
\end{align}
where $\omega$ and $\eta$ are the perturbation's frequency and steepness, determining its periodicity and localization, respectively.
The localized function with the sech form is set because it has the following limit value,
\begin{align}
    p_t^{(\pm)}=\lim_{t\rightarrow \pm \infty}p_t=i\omega\mp \eta.
\end{align}
By this approximation, the function $p_t(t,z)$ can be successfully transformed into a complex constant, and therefore offers great convenience for our analysis.
Moreover, this approximation can be conveniently applied to the function $p_z$.
Considering that the cases of $t\rightarrow \pm\infty$ have the same results, we take the case of $t\rightarrow -\infty$ as an example and can obtain
\begin{align}
    p_z^{(-)}=&\sqrt{M^{(-)}(2\beta-M^{(-)})},
\end{align}
where
\begin{align}
    M^{(-)}=&(i\omega+\eta)^2[\frac{\beta_2}{2}-\frac{\beta_4}{24}(i\omega+\eta)^2].
\end{align}
It describes the propagation characteristics of the perturbation.
To be specific, the perturbation's propagation constant and growing rate relative to the plane wave are respectively
\begin{align}
    K={\rm Im}[p_z^{(-)}],\quad G={\rm Re}[p_z^{(-)}].
\end{align}
Then, the velocity of its localized envelope and stripes are
\begin{align}
    V_{\rm en}=G/\eta,\quad V_{\rm st}=K/\omega.
\end{align}

From the insets of Fig. \ref{pic-ds}, we know that a SDS has a zero-velocity localized envelope and some zero-velocity stripes, which indicates $V_{\rm en}=0$ and $V_{\rm st}=0$.
In Fig. \ref{pic-mi} (a), when $\beta_2=-0.5$, $\beta_4=1$, and $\beta=1$, we show the curves of  $V_{\rm en}=0$ and $V_{\rm st}=0$ in the $\omega$-$\eta$ plane.
The coordinate of their intersection point $(\omega_s,\eta_s)$ is just the oscillation frequency and steepness of SDS, and the expressions are
\begin{align}\label{eq-omet-sds}
    \omega_s={\rm Re}[\sqrt{\frac{-3\beta_2+2\sqrt{3\beta\beta_4}}{\beta_4}}],\;
    \eta_s={\rm Re}[\sqrt{\frac{3\beta_2+2\sqrt{3\beta\beta_4}}{\beta_4}}].
\end{align}
Thus, the parameters need to satisfy $\omega_s\neq 0$ and  $\eta_s\neq 0$ for the existence of SDS, namely
\begin{align}\label{eq-ec-sds}
    \beta_4>{3\beta_2^2}/{4\beta},
\end{align}
as shown in Fig. \ref{pic-ph}.
Besides, one can find that they have the relationship,
\begin{align}
    \omega_s^2+\eta_s^2=\frac{4\sqrt{3\beta}}{\sqrt{\beta_4}},\quad \omega_s^2-\eta_s^2=-\frac{6\beta_2}{\beta_4}.
\end{align}
The former is independent of $\beta_2$, and the latter is independent of $\beta$.
Thus, the point $(\omega_s,\eta_s)$ always locates on a circle when $\beta_4$ and $\beta$ are fixed, and it moves towards the positive axis of $\eta$ with $\beta_2$ increasing, as shown in Fig. \ref{pic-mi} (a).
More interestingly, this point is just an exceptional point of the eigenvalue $p_z$, which is related to the non-Hermiticity of the matrix in Eq. (\ref{eq-abc}).
By calculating the contour integral of its argument, one can obtain that its winding number is $\pi$ \cite{Ding-2022}.
In eigenvalue problems, an exceptional point is a singularity where the two eigenvalues coalesce and each of them undergoes the transition from a real value to a complex.
With the development of non-Hermitian physics, the exceptional point plays an important role in various systems, like optics, photonics, and Bose–Einstein condensates \cite{Miri-2019,El-Ganainy-2018}.
However, for the relationship between SDS and exceptional point, a deeper understanding needs further investigations.
Next, to verify the effectiveness of our prediction, the dependence of $\omega_s$ on $\beta_2$ is studied.
When $\beta_4=1$ and $\beta=1$, we numerically measure the oscillation frequency of SDSs under different values of $\beta_2$ and show them by the red dots in Fig. \ref{pic-mi} (b).
One can see that $\omega_s$ gradually decreases into 0 with $\beta_2$ increasing.
To obtain them, we measure the time interval between the first and second peaks on the right side of SDS as its time period $T_s$, and then the oscillation frequency can be calculated by $\omega_s=2\pi /T_s$, as shown in the upper row of Fig. \ref{pic-ds} (a).
On the other hand, we also show the analytic prediction from the MLSA method by the black curve.
There are good agreements between the two results.

\begin{figure*}[htbp]
	\centering
	\includegraphics[width=176mm]{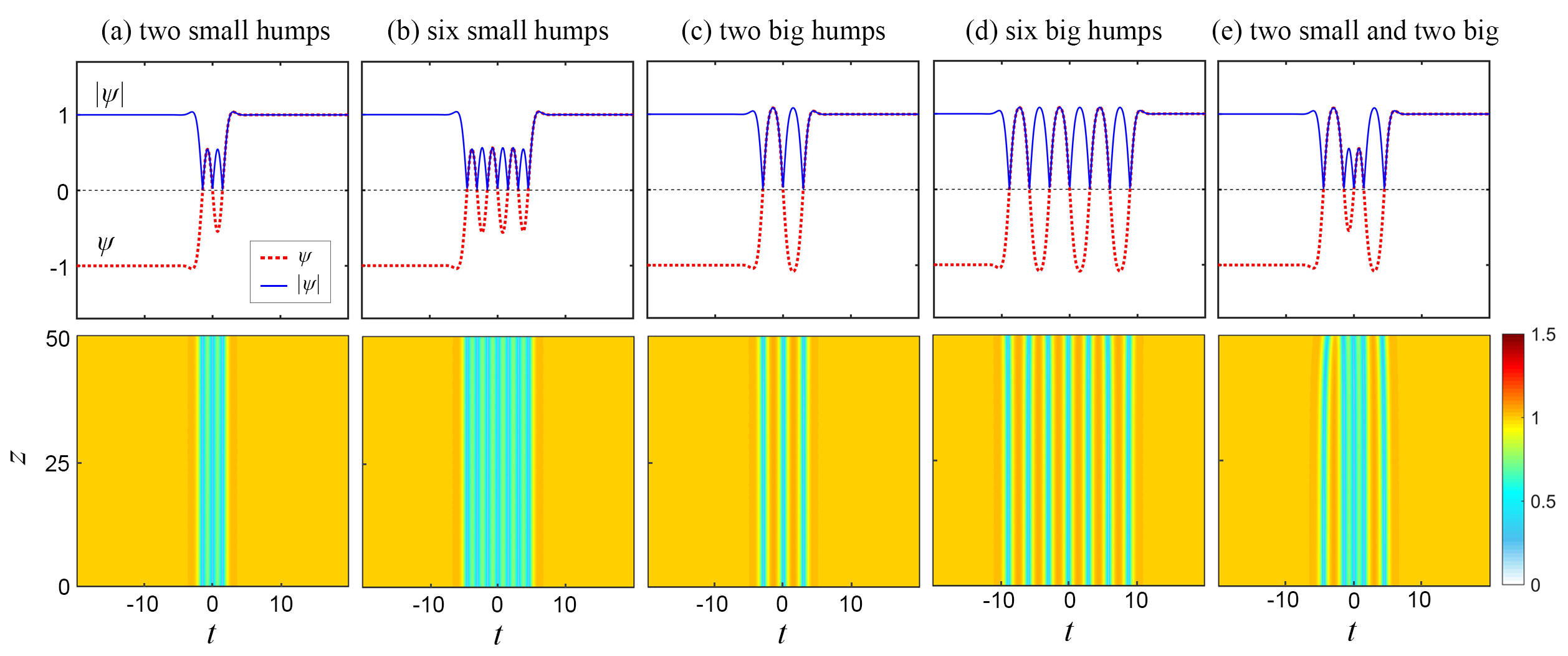}
	\caption{(Color online) The multi-dark-soliton bound states. The upper row is the distribution of their wave functions $\psi$ (red dashed curves) and amplitude $|\psi|$ (blue solid curves), when $\beta_2=0$, $\beta_4=1$, and $\beta=1$.
		The lower row is their corresponding amplitude evolution.
	}
	\label{pic-dsc}
\end{figure*}

\subsection{Analysis of propagation stability}

Another problem worthy of consideration is the propagation stability of SDSs.
From the insets of Fig. \ref{pic-ds}, one can see that some spontaneous oscillations emerge in the propagating process of SDSs.
It is a widespread phenomenon in the nonlinear systems with MI, and therefore prompts us to analyze the MI in the model (\ref{eq-model}).
As we know, different types of perturbations can produce different phenomena under the influence of MI.
Generally speaking, a purely periodic perturbation can generate Akhmediev breathers \cite{Dudley-2014,Dudley-2009}; a purely localized one can generate a rogue wave \cite{Zhao-2016-JOSAB} or Kuznetsov-Ma breather \cite{Duan-2019}, and induce the spontaneous oscillations in a triangular region \cite{El-1993,Biondini-2016}. (The word "purely" in front of "perturbation" is used to distinguish them with a periodic-localized perturbation.)
A SDS at the initial distance can be seen as a periodic-localized perturbation on a plane wave, so we need to separately consider the influence of MI on its periodic and localized components.
For its periodic component, we know that the traditional growing rate $g$ can be obtained by the modified growing rate $G$ when $\eta=0$, namely
\begin{align}
    g(\omega)&=|G(\omega,\eta=0)|\nonumber\\
    &=\Big|\omega\;{\rm Re}\Big[\sqrt{-(\frac{\beta_2}{2}+\frac{\beta_4}{24}\omega^2)(2\beta+\frac{\beta_2}{2}\omega^2+\frac{\beta_4}{24}\omega^4})\Big]\Big|.
\end{align}
If the SDS's oscillation frequency $\omega_s$ locates in the MI band, its periodicity will intensively impact the stability of plane wave, whose corresponding MI distribution is shown as the upper plot in Fig. \ref{pic-mi} (c).
Thus, we calculate the growing rate when $\omega=\omega_s$ by
\begin{align}\label{eq-goms}
    g(\omega_s)&=\frac{1}{8\beta_4}\sqrt{-\sqrt{3}\beta_2+2\sqrt{\beta\beta_4}}\nonumber\\
    &\times\sqrt{\sqrt{3}\beta_2(-9\beta_2^2+68\beta\beta_4)+2\sqrt{\beta\beta_4}(9\beta_2^2+20\beta\beta_4)},
\end{align}
to describe the MI growing rate induced by the periodicity of SDSs.
Because it can produce MI phenomena faster than the localization component, we call the periodicity-induced MI "strong MI", and accordingly we call the localization-induced MI "weak MI".
The typical distribution of growing rate of weak MI is shown as the lower plot in Fig. \ref{pic-mi} (c).
In Fig. \ref{pic-mi} (d), we illustrate the phase diagram of strong MI, weak MI, and modulation stability (MS) on the $\beta_2$-$\beta_4$ plane, and the growing rate of strong MI is presented by the color scale.
The strong MI corresponds to the region of  $\beta_2<0$ and $\frac{3\beta_2^2}{4\beta}<\beta_4<\frac{27\beta_2^2}{4\beta}$, and the weak MI is in the region of $\beta_2<0$ and $\beta_4>\frac{27\beta_2^2}{4\beta}$.
When $\beta_4=1$, the strong and weak MI respectively need the conditions of $-2/\sqrt{3}<\beta_2<-2/9$ and $\beta_2>-2/9$.
It is consistent with the evolution plots in Fig. \ref{pic-ds}, where the SDS is more and more stable with $\beta_2$ increasing.

\subsection{Multi-soliton bound state}

In this subsection, we focus on the multi-dark-soliton bound states.
In Ref. \cite{Buryak-1995}, the bound states of several bright solitons under the FOD have been found, and their existence condition and stability were analyzed.
Here, we will find the bound states of dark solitons by the biconjugate gradient method \cite{Press-book}.
%Considering that these states still have stripes, we call them "the SDSs with compound structures".
When $\beta_2=0$, $\beta_4=1$, and $\beta=1$, the distributions of wave functions of the multi-dark-soliton bound states are shown as the upper row in Fig. \ref{pic-dsc}.
There are many kinds of bound states, which have different number of amplitude humps, such as (a) two small humps, (b) six small humps, (c) two big humps, (d) six big humps, and (e) two small and two big humps.
The bound states include but are not limited to these five kinds, and their counterparts without the total phase step also exist but are not shown in this paper.
Next, we study their numerical evolution to test their propagation stability, which are shown as the lower row in Fig. \ref{pic-dsc}.
These bound states can always propagate for a long distance, and no MI phenomenon emerges.

We stress that the ``dispersion waves'' in Figs. \ref{pic-ds} (a-c) are the manifestation of the MI of plane-wave backgrounds, which is induced by the unstable frequency sideband from the periodicity of dark solitons. As shown in Fig. \ref{pic-mi} (a), the MI can be strong, weak, or absent in different cases, and therefore the dark solitons can propagate for a short or long distance under the influence of MI. Considering that they could be observed in the optical fibers with different length, our calculations about dark solitons can provide the theoretical guidance for their observations. Particularly, when $\beta_4\geq 0$, the dark solitons and their bound states are stable and propagate for a long distance, as shown in Figs. \ref{pic-ds} (d,e) and \ref{pic-dsc}.
In addition, the traditional linear stability analysis can also be used to obtain the linear equation (5) and the classical dispersion law (19). 
However, they can give us only the MI condition of a perturbed plane wave, but cannot give anything about striped solitons. 
This is the reason why we use the MLSA method to analyze the SDSs.

\begin{figure}[htbp]
	\centering
	\includegraphics[width=86mm]{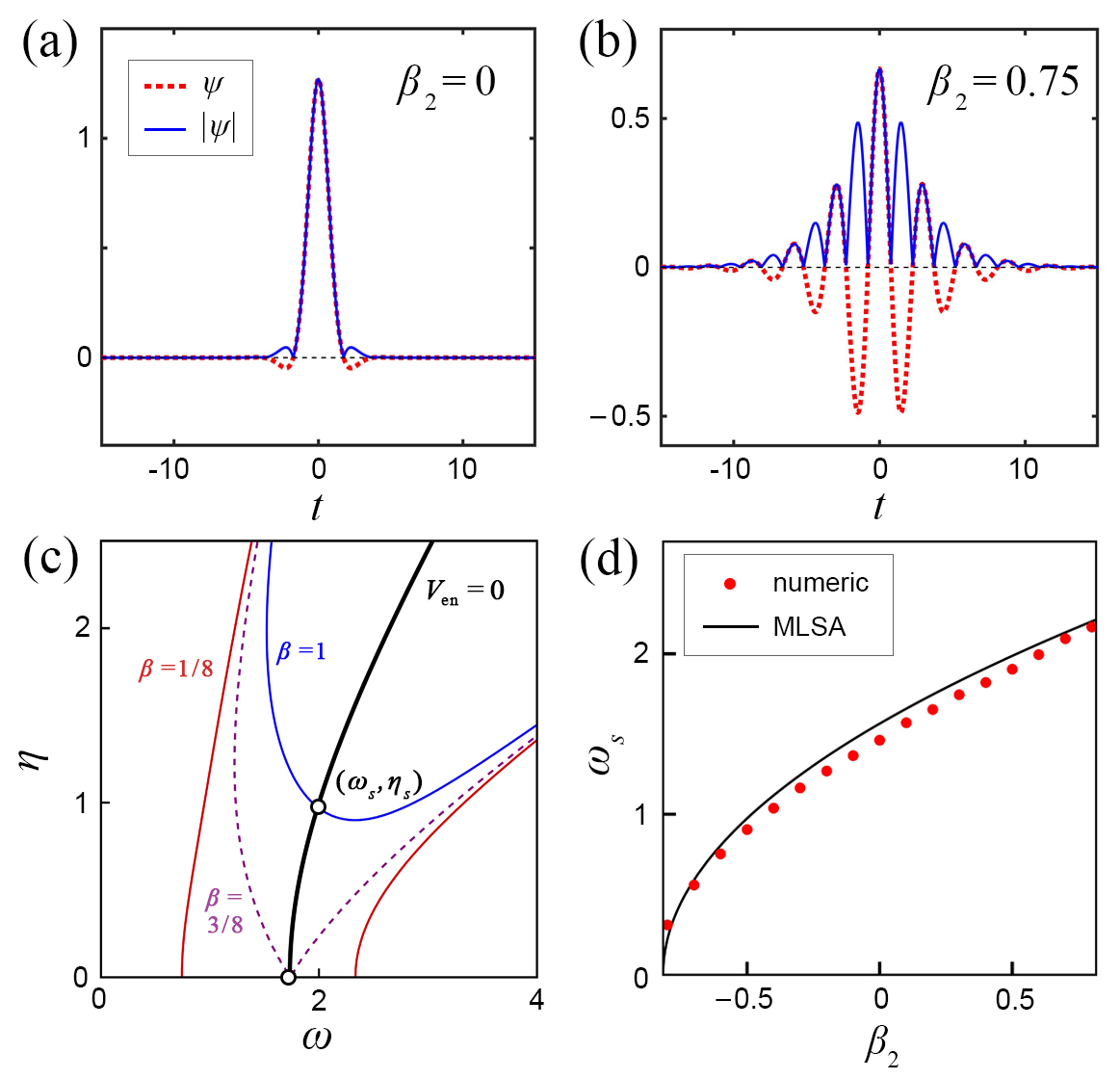}
	\caption{(Color online) (a) Distribution of the SBS's wave function $\psi$ (red dashed curves) and its amplitude $|\psi|$ (blue solid curves), when $\beta_2=0$, $\beta_4=-1$ and $\beta=1$. 
		(b) Same as (a) except for $\beta_2=0.75$.
		(c) In the $\omega$-$\eta$ plane, the characteristic parameters $(\omega_s, \eta_s)$ is the intersection point between the curves of $V_{\rm en}=0$ (black solid curve) and different $\beta$ value. The blue solid, purple dashed, and red solid curves denote $\beta=1$, $3/8$, and $1/8$, respectively. The parameters are $\beta_2=0.5$ and $\beta_4=-1$.
		(d) Dependence of SBS's oscillation frequency $\omega_s$ on $\beta_2$, when $\beta_4=1$ and $\beta=1$. The red dots and the black solid curve are respectively the results of numerical solutions and the MLSA method [see Eq. (\ref{eq-omet2})].
	}
	\label{pic-bs}
\end{figure}

\section{SBS}

As we know, the bright soliton with stripe structures in the model (\ref{eq-model}) has been studied in Refs. \cite{Zakharov-1998,Tam-2020}, including their existence condition and oscillation frequency.
Here, we will study them by the MLSA method and then provide a possible interpretation of their formation mechanism.

Firstly, by the biconjugate gradient method, we show the distribution of wave functions of SBS in Fig. \ref{pic-bs} (a) when $\beta_2=0$, $\beta_4=-1$, where the propagation constant is $\beta=1$. 
A bright soliton with oscillating tails can be seen, which is just the so-called pure-quartic soliton \cite{Blanco-2016}.
Then, when we change $\beta_2$ to $0.75$, a bright soliton with many stripes appears in Fig. \ref{pic-bs} (b).
Unlike the SDS, the SBS has on background of plane wave.
Thus, our calculation of MLSA directly starts with the perturbation,
\begin{align}\label{eq-psiap}
    \psi_p=u=A e^{p(t,z)}+B e^{p^*(t,z)},
\end{align}
where the function $p(t,z)$ still has the form of Eq. (\ref{eq-p}) at the initial distance.
Then, substituting Eq. (\ref{eq-psiap}) into the model (\ref{eq-model}) and considering that the perturbation has a small amplitude, one can obtain
\begin{align}
    p_z=&i\{p_t^2[-\frac{\beta_2}{2}+\frac{\beta_4}{24}(p_t^2+p_{tt})]\nonumber\\
    &+p_{tt}(-\frac{\beta_2}{2}+\frac{\beta_4}{8}p_{tt})+\frac{\beta_4}{6}p_tp_{ttt}+\frac{\beta_4}{24}p_{tttt}\}.
\end{align}
After taking the limitations of $t\rightarrow -\infty$, the propagation function becomes
\begin{align}
    p_z^{(-)}=i(i\omega+\eta)^2[-\frac{\beta_2}{2}+\frac{\beta_4}{24}(i\omega+\eta)^2].
\end{align}
It indicates that the propagation constant of this wave is
\begin{align}\label{eq-beta}
    \beta={\rm Im} [p_z^{(-)}]=-\frac{\beta_2}{2}(\eta^2-\omega^2)+\frac{\beta_4}{24}[(\eta^2-\omega^2)^2-4\omega^2\eta^2].
\end{align}
It is worth noting that the formation mechanism of stripes in SDS and SBS is different.
For SDS, the stripes originate from the interaction between a perturbation and a zero-velocity plane wave; for SBS, the stripes are caused by the interaction between two components with opposite frequencies.
Thus, the velocity of stripes in SBS is not equal to $V_{\rm st}=K/\omega$.
Nevertheless, the formation of localized envelopes is not related to the background wave, so $V_{\rm en}$ can still describe the velocity of localized envelopes in SBS.
Its expression is 
\begin{align}
    V_{\rm en}=G/\eta={\rm Re}[p_z^{(-)}]/\eta=\omega[\beta_2+\frac{\beta_4}{6}(\omega^2-\eta^2)].
\end{align}

For a certain propagation constant $\beta$, by solving $V_{\rm en}=0$ and Eq. (\ref{eq-beta}), we can obtain that the oscillation frequency and steepness of SBS,
\begin{align}\label{eq-omet2}
    \omega_s={\rm Re}[\sqrt{\frac{3\beta_2+\sqrt{-6\beta\beta_4}}{-\beta_4}}],\;
    \eta_s={\rm Re}[\sqrt{\frac{-3\beta_2+\sqrt{-6\beta\beta_4}}{-\beta_4}}].
\end{align}
As shown in Fig. \ref{pic-bs} (c), the point $(\omega_s, \eta_s)$ is the intersection of the black curve ($V_{\rm en}=0$) and the blue curve ($\beta=1$), when $\beta_2=0.5$ and $\beta_4=-1$.
We also depict the curves of $\beta=3/8$ and $\beta=1/8$: the former is the critical case where the intersection exists just right, and the latter has no intersection with the curve of $V_{\rm en}=0$.
It indicates that the SBS cannot exist when $\beta<3/8$, and we also obtain the same result in our numerical simulations.
Then, considering that $\omega_s\neq 0$ and  $\eta_s\neq 0$, we can obtain that the existence condition of SBS is
\begin{align}\label{eq-ec-sbs}
    \beta_4<-{3\beta_2^2}/{2\beta},
\end{align}
as shown in Fig. \ref{pic-ph}.
We also show the oscillation frequency of SDS under different sets of $\beta_2$ in Fig. \ref{pic-bs} (d), when $\beta_4=1$ and $\beta=1$. 
Our analytical prediction (\ref{eq-omet2}) shows a good agreement with the numerical result.

\begin{figure}[htbp]
	\centering
	\includegraphics[width=86mm]{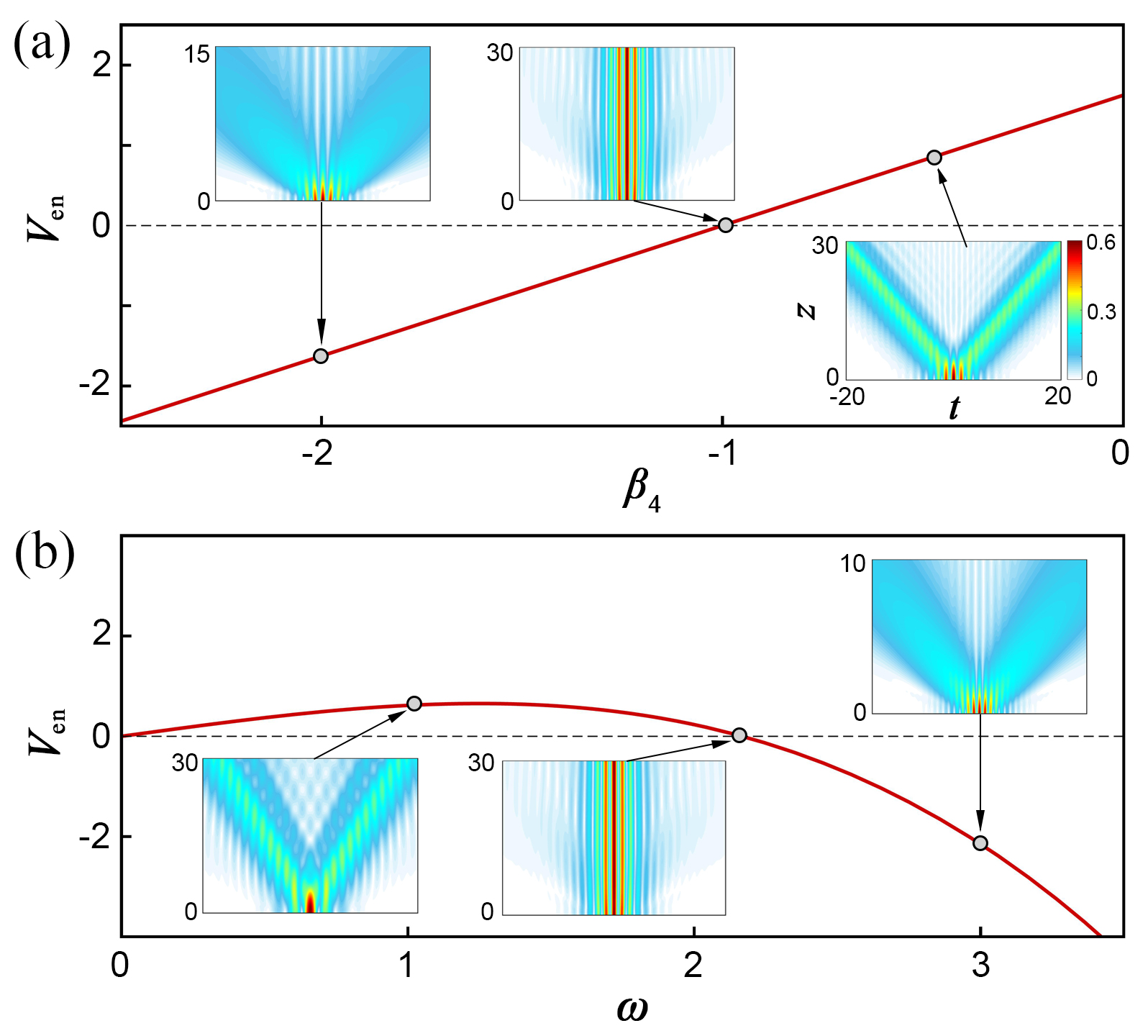}
	\caption{(Color online) (a) Dependence of $V_{\rm en}$ on $\beta_4$ when $\omega=2.1678$, $\eta=0.4466$, and $\beta_2=0.75$. The insets are the amplitude evolution of the initial state (\ref{eq-bsini}) when $\beta_4=-2$, $-1$, and $-0.5$.
		(b) Dependence of $V_{\rm en}$ on $\omega$ when $\eta=0.4466$, $\beta_2=0.75$, and $\beta_4=-1$. The insets are the amplitude evolution of the initial state (\ref{eq-bsini}) when $\omega=1$, $2.1678$, and $3$.
	}
	\label{pic-bsv}
\end{figure}

\renewcommand\arraystretch{1.5}
\begin{table*}[htbp]
\caption{Characteristics of several nonlinear waves}
\label{tab-1}
\begin{tabular}{p{4.2cm}<{\centering}|p{2.5cm}<{\centering}|p{2.5cm}<{\centering}|p{2.5cm}<{\centering}|p{2.5cm}<{\centering}}
	\hline
	\hline
	& $\;\;$ Periodicity \newline ($\omega_s$) & $\;\;$ Localization \newline ($\eta_s$) & Proportion ($\omega_s/\eta_s$) & Characteristic quantity ($\varphi$)\\
	\hline
	Traditional soliton & No ($\omega_s=0$) & Yes ($\eta_s> 0$) & 0 & $0$\\
	\hline
	Striped soliton & Yes ($\omega_s>0$) & Yes ($\eta_s> 0$) & $(0,+\infty)$ & $(0,{\pi}/{2})$\\
	\hline
	Pure-quartic soliton & Yes ($\omega_s>0$) & Yes ($\eta_s> 0$) & $1$ & ${\pi}/{4}$\\
	\hline
	Periodic wave & Yes ($\omega_s>0$) & No ($\eta_s= 0$) & $+\infty$ & ${\pi}/{2}$\\
	\hline
	\hline
\end{tabular}
\end{table*}

Our derived existence condition (\ref{eq-ec-sbs}) and oscillation frequency (\ref{eq-omet2}) of SBSs agree well with the results given by Ref. \cite{Tam-2020}.
It indicates that for SBSs, the MLSA method are equivalent to the direct asymptotic analysis, due to the absence of plane-wave background.
Nevertheless, besides of the above results, the MLSA method can also provide the interpretation for the formation mechanism of SBSs.
It is known that the wave function (\ref{eq-psiap}) has two components with opposite frequencies.
When both of their localized envelopes have the zero velocity, i.e., $V_{\rm en}=0$, a SBS will be generated.
It means that the formation of a SBS is the result of synchronous propagation of the two components.
To confirm this guess, for a wave function with frequency $\omega$ and steepness $\eta$, we show the dependence of $V_{\rm en}$ on $\beta_4$ in Fig. \ref{pic-bsv} (a), when $\omega=2.1678$, $\eta=0.4466$, and $\beta_2=0.75$.
When $\beta_4=-1$, we have $V_{\rm en}=0$.
The initial state has the form of
\begin{align}\label{eq-bsini}
    \psi(t,0)=a \cos(\omega t){\rm sech}(\eta t),
\end{align}
where the amplitude $a$ is set as $0.6$ according to the numerical solution of the SBS in this case.
Its amplitude evolution plot is shown in the middle inset of Fig. \ref{pic-bsv} (a), where a SBS is generated.
When $\beta_4=-2$, we have $V_{\rm en}<0$.
Its corresponding evolution plot is shown in the left insets of Fig. \ref{pic-bsv} (a), where the two components move towards opposite directions.
Also, the similar result can be obtained in the case of $\beta_4=-0.5$, as shown in the right inset.
It indicates that one cannot find SBS in the absence of FOD, because the velocity of localized envelopes is not zero in this case.

Similarly, we also show the dependence of $V_{\rm en}$ on $\omega$ in Fig. \ref{pic-bsv} (b), when $\eta=0.4466$, $\beta_2=0.75$, and $\beta_4=-1$.
The evolution plots in the cases of $\omega=1$, $2.1678$, and $3$ are shown as the insets in Fig. \ref{pic-bsv}.
Only in the case of $\omega=2.1678$, the SBS can be generated. 
In the cases of $\omega\neq 2.1678$, the two components propagate towards opposite directions.

Finally, we summarize the characteristics of different nonlinear waves in Tab. \ref{tab-1}.
They contains the traditional solitons, the striped solitons, the pure-quartic solitons, and periodic waves.
We recall that the quantity $\varphi$ is defined by $\varphi=\tan^{-1} ({\omega_s}/{\eta_s})$ to describe the strength of a wave's periodicity relative to its localization, and $\omega_s$ and $\eta_s$ are the parameters determining the periodicity and localization of a wave, respectively.
A traditional soliton has the localization but no the periodicity (i.e., $\omega_s=0$ and $\eta_s>0$), while a periodic wave has the periodicity but no the localization (i.e., $\omega_s>0$ and $\eta_s=0$).
A striped soliton has both of the periodicity and localization, and therefore it corresponds to $\omega_s>0$ and $\eta_s>0$.
The pure-quartic soliton is a particular case of the striped solitons when the periodicity and localization achieve a balance, i.e., $\omega_s=\eta_s$, so its proportion of periodicity and localization is $\omega_s/\eta_s=1$ and the characteristic quantity is $\varphi=\pi/4$.
\\

\section{Conclusion}

In summary, the excitations of dark solitons in a nonlinear fiber with the second-order dispersion and FOD are investigated.
In this system, we find the SDSs, their counterparts without the total phase step, and some multi-dark-soliton bound states.
With the help of the MLSA method, we quantitatively analyze the periodicity and localization of the SDSs under different coefficients of second-order dispersion and FOD.
Their oscillation frequency is successfully predicted and the corresponding MI is analyzed, which can make the propagation of SDSs unstable.
For the SBSs, though their major characteristics have been studied in the previous works \cite{Zakharov-1998,Tam-2020}, we provide a new possible perspective to understand their formation mechanism.
Our method and results can provide the theoretical guidance for the generations and manipulations of striped solitons in nonlinear fibers, and furthermore bring more possibility for their applications in soliton communications and optical measurements.
Considering that stripe solitons have been also found in spinor Bose-Einstein condensates \cite{Achilleos-2013,Xu-2013,Zhao-2020,Yang-2021}, we expect that our method can be applied in these systems and help with the observation of matter-wave striped solitons.
We also note that the competition between periodicity and localization of a soliton is similar to the well-known wave-particle duality to some extent.
Understanding the similarity between them is a challenging but might be important task, which is worthy of future’s study.

\section*{Acknowledgement}
The authors thank Prof. Jie Liu for his helpful discussion.
This work was supported by NSAF (No.U2330401) and National Natural Science Foundation of China (No. 12247110).


\begin{thebibliography}{}

\bibitem{Hecht-book}  E. Hecht, Optics (Addison-Wesley, Reading, MA, 2001).
\bibitem{Cronin-2009} A. D. Cronin, J. Schmiedmayer, and D. E. Pritchard, Optics and interferometry with atoms and molecules, Rev. Mod. Phys. 81, 1051 (2009).
%\bibitem{Wright-2023} A. Wright, A century of matter waves, Nat. Rev. Phys. 5, 318 (2023).
\bibitem{Kivshar-book} Y. S. Kivshar and G. P. Agrawal, Optical solitons: from fibers to photonic crystals (Academic Press, 2003).
\bibitem{Onorato-2013} M. Onorato, S. Residori, U. Bortolozzo, A. Montina, and F. T. Arecchi, Rogue waves and their generating mechanisms in different physical contexts, Phys. Rep. 528, 47 (2013).
\bibitem{Dudley-2014} J. M. Dudley, F. Dias, M. Erkintalo, and G. Genty, Instabilities, breathers and rogue waves in optics, Nat. Photonics 8, 755 (2014).
\bibitem{Mollenauer-1980} L. F. Mollenauer, R. H. Stolen, and J. P. Gordon, Experimental observation of picosecond pulse narrowing and solitons in optical fibers, Phys. Rev. Lett. 45, 1095 (1980).
\bibitem{Weiner-1988} A. M. Weiner, J. P. Heritage, R. J. Hawkins, R. N. Thurston, E. M. Kirschner, D. E. Leaird, and W. J. Tomlinson, Experimental observation of the fundamental dark soliton in optical fibers, Phys. Rev. Lett. 61, 2445 (1988).
\bibitem{Agrawal-book} G. P. Agrawal, Nonlinear fiber optics (Academic Press, New York, 2007).
\bibitem{Kevrekidis-book} P. G. Kevrekidis, D. J. Frantzeskakis, and R. Carretero-González, Emergent nonlinear phenomena in Bose-Einstein condensates: Theory and experiment (Springer, New York, 2008).
\bibitem{Haus-1996} H. A. Haus and W. S. Wong, Solitons in optical communications, Rev. Mod. Phys. 68, 423 (1996).
\bibitem{McDonald-2014} G. D. McDonald, C. C. N. Kuhn, K. S. Hardman, S. Bennetts, P. J. Everitt, P. A. Altin, J. E. Debs, J. D. Close, and N. P. Robins, Bright solitonic matter-Wave interferometer, Phys. Rev. Lett. 113, 013002 (2014).
%\bibitem{Helm-2015} J. L. Helm, S. L. Cornish, and S. A. Gardiner, Sagnac interferometry using bright matter-Wave solitons, Phys. Rev. Lett. 114, 134101 (2015).
\bibitem{Snyder-1997} A. W. Snyder and D. J. Mitchell, Accessible solitons, Science 276, 1538 (1997).
\bibitem{Kumar-2009} V. R. Kumar, R. Radha, and P. K. Panigrahi, Matter wave interference pattern in the collision of bright solitons. Phys. Lett. A 373, 4381 (2009).
\bibitem{Zhao-2016} L. C. Zhao, L. Ling, Z. Y. Yang, and J. Liu, Properties of the temporal–spatial interference pattern during soliton interaction, Nonlinear Dynam. 83, 659 (2016).
\bibitem{Xiong-2023} W. Xiong, P. Gao, Z. Y. Yang, and W. L. Yang, Quantized reflection of a soliton by a vibrating atomic mirror, Phys. Rev. A 108, 023303 (2023).
\bibitem{Gao-2023} P. Gao and J. Liu, Quantum scattering treatment on the time-domain diffraction of a matter-wave soliton, Phys. Rev. A 109, 013323 (2024).
\bibitem{Zabusky-1965} N. J. Zabusky and M. D. Kruskal, Interaction of “solitons” in a collisionless plasma and the rence of initial states, Phys. Rev. Lett. 15, 240 (1965).
\bibitem{Stegeman-1999} G. I. Stegeman and M. Segev, Optical spatial solitons and their interactions: Universality and diversity, Science 286, 1518 (1999).
%\bibitem{Martin-2007} A. D. Martin, C. S. Adams, and S. A. Gardiner, Bright matter-wave soliton collisions in a harmonic trap: Regular and chaotic dynamics, Phys. Rev. Lett. 98, 020402 (2007).
\bibitem{Akhmediev-1994} N. N. Akhmediev, A. V. Buryak, and M. Karlsson, Radiationless optical solitons with oscillating tails, Opt. Commun. 110, 540 (1994).
\bibitem{Buryak-1995} A. V. Buryak and N. N. Akhmediev, Stability criterion for stationary bound states of solitons with radiationless oscillating tails, Phys. Rev. E 51, 3572 (1995).
\bibitem{Blanco-2016} A. Blanco-Redondo, C. Martijn de Sterke, J. E. Sipe, T. F. Krauss, B. J. Eggleton, and C. Husko, Pure-quartic solitons, Nat. Commun. 7, 10427 (2016).
\bibitem{Tam-2019} K. K. K. Tam, T. J. Alexander, A. Blanco-Redondo, and C. M. de Sterke, Stationary and dynamical properties of pure-quartic solitons, Opt. Lett. 44, 3306 (2019).
\bibitem{Zakharov-1998} V. E. Zakharov and E. A. Kuznetsov, Optical solitons and quasisolitons, J. Exp. Theor. Phys. 86, 1035 (1998).
\bibitem{Tam-2020} K. K. K. Tam, T. J. Alexander, A. Blanco-Redondo, and C. M. de Sterke, Generalized dispersion Kerr solitons, Phys. Rev. A 101, 043822 (2020).
\bibitem{Gao-2020} P. Gao, C. Liu, L. C. Zhao, Z. Y. Yang, and W. L. Yang, Modified linear stability analysis for quantitative dynamics of a perturbed plane wave, Phys. Rev. E 102 022207 (2020).
\bibitem{Gao-2021} P. Gao, L. Duan, X. Yao, Z. Y. Yang, and W. L. Yang, Controllable generation of several nonlinear waves in optical fibers with third-order dispersion, Phys. Rev. A 103 023519 (2021).
\bibitem{Runge-2020} A. F. J. Runge, D. D. Hudson, K. K. K. Tam, C. M. de Sterke, and A. Blanco-Redondo, The pure-quartic soliton laser, Nat. Photonics 14, 492 (2020).

\bibitem{Karlsson-1994} M. Karlsson and A. Höök, Soliton-like pulses governed by fourth order dispersion in optical fibers, Opt. Commun. 104, 303 (1994).
\bibitem{Piche-1996} M. Piché, J.-F. Cormier, and X. Zhu, Bright optical soliton in the presence of fourth-order dispersion, Opt. Lett. 21, 845 (1996).

\bibitem{Press-book} W. H. Press, Numerical Recipes (Cambridge University Press, Cambridge, 1990).
\bibitem{Winiecki-1999} T. Winiecki, J. F. McCann, and C. S. Adams, Vortex structures in dilute quantum fluids, Europhys. Lett. 48, 475 (1999).
\bibitem{Edmonds-2016} M. J. Edmonds, T. Bland, D. H. J. O'Dell, and N. G. Parker, Exploring the stability and dynamics of dipolar matter-wave dark solitons, Phys. Rev. A 93, 063617 (2016).
\bibitem{Yang-book} J. Yang, Nonlinear Waves in Integrable and Nonintegrable Systems (Siam, Philadelphia, 2010).

\bibitem{Ding-2022} K. Ding, C. Fang, and G. Ma. Non-Hermitian topology and exceptional-point geometries. Nat. Rev. Phys. 4, 745 (2022).
\bibitem{El-Ganainy-2018} R. El-Ganainy, K. G. Makris, M. Khajavikhan, Z. H. Musslimani, S. Rotter, and D. N. Christodoulides, Non-Hermitian physics and PT symmetry, Nat. Phys. 14, 11–19 (2018).
\bibitem{Miri-2019} M. -A. Miri and A. Alù, Exceptional points in optics and photonics, Science 363, eaar7709 (2019).

\bibitem{Dudley-2009} J. M. Dudley, G. Genty, F. Dias, B. Kibler, and N. Akhmediev, Modulation instability, Akhmediev Breathers and continuous wave supercontinuum generation, Opt. Express 17, 21497 (2009).
\bibitem{Zhao-2016-JOSAB} L. C. Zhao and L. Ling, Quantitative relations between modulational instability and several well-known nonlinear excitations, J. Opt. Soc. Am. B 33, 850 (2016).
\bibitem{Duan-2019} L. Duan, Z. Y. Yang, P. Gao, and W. L. Yang, Excitation conditions of several fundamental nonlinear waves on continuous-wave background, Phys. Rev. E 99, 012216 (2019).
\bibitem{El-1993} G. El, A. Gurevich, V. Khodorovskii, and A. Krylov, Modulational instability and formation of a nonlinear oscillatory structure in a “focusing” medium, Phys. Lett. A 177, 357 (1993).
\bibitem{Biondini-2016} G. Biondini and D. Mantzavinos, Universal nature of the nonlinear stage of modulational instability, Phys. Rev. Lett. 116, 043902 (2016).

\bibitem{Achilleos-2013} V. Achilleos, D. J. Frantzeskakis, P. G. Kevrekidis, and D. E. Pelinovsky, Matter-wave bright solitons in spin-orbit coupled Bose-Einstein condensates, Phys. Rev. Lett. 110, 264101 (2013).
\bibitem{Xu-2013} Y. Xu, Y. Zhang, and B. Wu, Bright solitons in spin-orbit coupled Bose-Einstein condensates, Phys. Rev. A 87, 013614 (2013).
\bibitem{Zhao-2020} L. C. Zhao, X. W. Luo, and C. Zhang, Magnetic stripe soliton and localized stripe wave in spin-1 Bose-Einstein condensates, Phys. Rev. A 101, 023621 (2020).
\bibitem{Yang-2021} Y. X. Yang, P. Gao, Z. Wu, L. C. Zhao, and Z. Y. Yang, Matter-wave stripe solitons induced by helicoidal spin–orbit coupling,
Ann. Phys. 431 168562 (2021).

\end{thebibliography}
\end{document}